\documentclass[superscriptaddress,prd,aps,showpacs,nofootinbib,showkeywords,eqsecnum]{revtex4-2}
\usepackage{graphicx,color,amsmath,amsxtra}
\usepackage{epsf}
\usepackage{amssymb}
\usepackage{enumerate}
\usepackage{hhline}
\usepackage{array}
\usepackage{tabularx}
\usepackage[unicode]{hyperref}             
\usepackage{epstopdf}
\begin{document}
\pagestyle{myheadings}
\title{{\it k}-Gauss-Bonnet inflation}
\author{Tuyen M. Pham}
\email{tuyen.phammanh@phenikaa-uni.edu.vn}
\affiliation{Phenikaa Institute for Advanced Study, Phenikaa University, Hanoi 12116, Vietnam}
\affiliation{Graduate University of Science and Technology, Vietnam Academy of Science and Technology, Hanoi 11307, Vietnam}
\author{Duy H. Nguyen}
\email{duy.nguyenhoang@phenikaa-uni.edu.vn}
\affiliation{Phenikaa Institute for Advanced Study, Phenikaa University, Hanoi 12116, Vietnam}
\affiliation{Graduate University of Science and Technology, Vietnam Academy of Science and Technology, Hanoi 11307, Vietnam}
\author{Tuan Q. Do }
\email{tuan.doquoc@phenikaa-uni.edu.vn}
\affiliation{Phenikaa Institute for Advanced Study, Phenikaa University, Hanoi 12116, Vietnam}
\affiliation{Faculty of Basic Sciences, Phenikaa University, Hanoi 12116, Vietnam}
\date{\today} 
\begin{abstract}
We propose a novel {\it k}-Gauss-Bonnet inflationary model, in which a kinetic term of scalar field is allowed to non-minimally couple to the Gauss-Bonnet topological invariant in the absence of a potential of scalar field. As a result, this model is shown to admit an isotropic power-law inflation provided that the scalar field is phantom. Furthermore, stability analysis based on the dynamical system method is performed to indicate that this inflation solution is indeed stable and attractive. More interestingly, a gradient instability in tensor perturbations is shown to disappear in this model.
%Keywords: expanding universes, Bianchi spaces, no-hair theorem, CMB
\end{abstract}
\maketitle
%%%%%%%%%%%%%%%%%%%%%%%%%%%%%%%%%%%%%%%%%%
\section{Introduction}
Inflationary universe \cite{Starobinsky:1980te,Guth:1980zm,Linde:1981mu,Linde:1983gd}, which was firstly introduced as a solution to the puzzles of the standard hot big bang cosmology such as the flatness, horizon and magnetic-monopole ones by Guth,  has been regarded as one of the leading paradigms in modern cosmology due to the fact that many of its theoretical predictions have been well confirmed by  the recent cosmic microwave background radiation (CMB) observations of the Wilkinson Microwave Anisotropy Probe (WMAP) and Planck  satellites \cite{Hinshaw:2012aka,Aghanim:2018eyx,Akrami:2018odb}. It turns out that there have been some potential approaches to realize the mechanism of inflation. The first approach, which is the dominant one, is based on the introduction of a hypothesis scalar field called an inflaton field. In this scenario, the scalar field is usually assumed to roll slowly down to its potential during an inflationary phase. Hence, the existence of the potential of scalar field is necessary and is required to be nearly flat. As a result, a huge number of scalar field models have been proposed to realize the origin of inflaton field along with its potential \cite{Martin:2013tda}. However, it is worth noting that there has been an exotic model called the {\it k}-inflation, in which inflation can be achieved even when the pure potential of scalar field is absent \cite{ArmendarizPicon:1999rj,Garriga:1999vw,Do:2020hjf}. More interestingly, some models which share the same idea of the inflationary universe driven by a non-minimal kinetic term with the {\it k}-inflation model have been constructed such as the ghost inflation \cite{ArkaniHamed:2003uz} or Galileon inflation \cite{Deffayet:2009wt,Deffayet:2010qz,Kobayashi:2010cm,Kobayashi:2011nu,Horndeski:1974wa,Do:2017qyd}. Among the other ones, the approach based on the higher-order modification of Einstein gravity such as the Starobinsky model involving the $R^2$ correction \cite{Starobinsky:1980te} or the  Gauss-Bonnet models \cite{Satoh:2007gn,Satoh:2008ck,Guo:2009uk,Guo:2010jr,Kanti:2015pda,Kanti:2015dra,Chakraborty:2018scm,Odintsov:2018zhw,Odintsov:2021lum}  involving the topological invariant, $G \equiv R^2 - 4R_{\mu\nu}R^{\mu\nu}+R_{\mu\nu\rho\sigma} R^{\mu\nu\rho\sigma}$ \cite{Lovelock:1971yv}, which appears as an $\alpha'$-order correction in the heterotic superstring  effective action \cite{Zwiebach:1985uq,Boulware:1985wk,Gross:1986mw,Metsaev:1987zx}, has played as one of the most promising approaches. Remarkably, the Starobinsky model has remained as one of the most favorable models in the light of the Planck observation \cite{Akrami:2018odb}, while the Gauss-Bonnet models have been extensively used to deal with other leading cosmological issues such as singularity-free universes \cite{Antoniadis:1993jc,Rizos:1993rt}, dark energy \cite{Nojiri:2005vv,Nojiri:2005jg}, and black holes \cite{Mignemi:1992nt,Kanti:1995vq,Kleihaus:2011tg,Antoniou:2017acq,Doneva:2017bvd,Silva:2017uqg}. It should be noted that the Gauss-Bonnet invariant acts as a total derivative in four dimensional spacetimes and therefore contributes nothing to the gravitational dynamics. In higher dimensional spacetimes, however, this property will no longer be valid. In four dimensional spacetimes, therefore, ones must couple the Gauss-Bonnet invariant to other field(s), usually  to scalar field(s) such as $f(\phi)G$ \cite{Satoh:2007gn,Satoh:2008ck,Guo:2009uk,Guo:2010jr,Kanti:2015pda,Kanti:2015dra,Chakraborty:2018scm,Odintsov:2018zhw,Odintsov:2021lum,Antoniadis:1993jc,Rizos:1993rt,Nojiri:2005vv,Nojiri:2005jg,Mignemi:1992nt,Kanti:1995vq,Kleihaus:2011tg,Antoniou:2017acq,Doneva:2017bvd,Silva:2017uqg}, in order to count its effect on the gravitational dynamics. This is due to the fact that the coupling $f(\phi)G$, where $f(\phi)$ is an arbitrary non-constant function of scalar field $\phi$, will no longer be a total derivative in four dimensional spacetimes. Interestingly, by regarding this coupling $f(\phi)G$ as an effective potential of scalar field, ones have claimed in Refs. \cite{Kanti:2015pda,Kanti:2015dra} that inflation can happen within a scalar-Gauss-Bonnet model in the absence of the pure potential of scalar field, i.e., $V(\phi)$. For additional discussions on this model, see Ref. \cite{Chakraborty:2018scm}. More interestingly, novel hairy black holes have been found in this model \cite{Antoniou:2017acq,Doneva:2017bvd,Silva:2017uqg}. Unfortunately, inflationary universe in this  model has been shown to be inviable due to its gradient instability in the tensor perturbations \cite{Hikmawan:2015rze}.  Recently,  a study in Ref. \cite{Glavan:2019inb} has claimed that the total derivative property of the Gauss-Bonnet invariant can be overcome by rescaling a coupling constant of the Gauss-Bonnet term as $\alpha \to \alpha/(D-4)$ then taking a limit $D\to 4$, where $D$ is the number of dimensions of spacetimes. Consequently, we might no longer need the coupling $f(\phi)G$. This idea has, therefore, received a lot of attention. Remarkably, some critical arguments to the validity of this idea have been drawn in the follow-up papers \cite{Gurses:2020rxb,Gurses:2020ofy,Arrechea:2020gjw}. As a result, it turns out that a non-trivial coupling between scalar field(s) and the Gauss-Bonnet invariant seems to be an essential part of four dimensional Gauss-Bonnet gravity. 

Inspired by the {\it k}-inflation \cite{ArmendarizPicon:1999rj,Garriga:1999vw} and the Gauss-Bonnet inflation \cite{Kanti:2015pda,Kanti:2015dra}, we would like to propose in this paper a novel scenario, in which the Gauss-Bonnet invariant is allowed to non-minimally couple to the kinetic term of scalar field $\phi$ in the absence of the pure potential $V(\phi)$. For convenience, we call it a {\it k}-Gauss-Bonnet (or {\it k}GB for short) gravity model. As a result, this model will be shown to admit an isotropic power-law inflation as its stable and attractive solution.
%%%%%%%%%%%%%%%%%%%
\section{Basic setup}
As a result, a general action of the {\it k}-Gauss-Bonnet gravity model is given by
\begin{align}
	S= \int{d^4 x \sqrt{-g}} \left[\frac{1}{2}R +P(X,\phi)-\frac{1}{8} J(X,\phi) G \right],
\end{align}
where the reduced Planck mass $M_p$ has been set to be one for convenience. In addition, $X\equiv - {\omega}\partial_\mu \phi \partial^\mu \phi/{2} $ is the kinetic term of scalar field $\phi$. It is noted that $\omega$ can be either $+1$ or $-1$ \cite{Guo:2009uk,Guo:2010jr}. Furthermore, $\omega =-1$ corresponds to a phantom scalar field \cite{Caldwell:1999ew,Do:2011zza}. Generally, $P(X,\phi)$ and $J(X,\phi)$ are arbitrary functions of $\phi$ and $X$.  In the rest of the paper, however, we will restrict ourselves to a simple scenario \footnote{Remarkably, we have just found two recent interesting papers \cite{Ghalee:2018qeo,Kaczmarek:2020awp}, which also consider the novel coupling between the kinetic of scalar field and the Gauss-Bonnet invariant. However, the first paper \cite{Ghalee:2018qeo} deals with the condensed state of scalar field and its implication for the late-time universe; while the latter one \cite{Kaczmarek:2020awp} studies the energy conditions in the Friedmann-Lemaitre-Robertson-Walker (FLRW) spacetime. Both of them do not investigate an inflationary universe. In addition, the phantom field is not considered in these two papers.}, in which both $P(X,\phi)$ and $J(X,\phi)$ depend only on the kinetic term of scalar field such as  $P(X,\phi)=X$ and $J(X,\phi)=J(X)$, i.e., 
\begin{align}\label{2.1}
	S= \int{d^4 x \sqrt{-g}} \left[\frac{1}{2}R +X-\frac{1}{8} J(X) G \right].
\end{align}
Apparently, this action has a shift symmetry, $\phi \to \phi + \text{const}$, similar to that of the Galileon inflation models \cite{Kobayashi:2010cm,Kobayashi:2011nu}. It should be noted that  Ref. \cite{Odintsov:2021lum} has proposed a scenario, in which a {\it k}-inflation correction is introduced to the usual Gauss-Bonnet gravity having both $V(\phi)$ and $f(\phi)G$. 

As a result, varying the action \eqref{2.1} with respect to the inverse metric $g^{\mu\nu}$ will lead to the corresponding Einstein field equation,
\begin{align}\label{2.3}
	&G_{\mu\nu}- \left(R_{\mu\rho\nu\sigma}-g_{\mu\nu}R_{\rho\sigma} \right)\nabla^{\rho}\nabla^{\sigma}J+G_{\mu\nu}\square J -R_{\nu\rho}\nabla_{\mu}\nabla^\rho J-R_{\mu\rho}\nabla_{\nu}\nabla^{\rho}J+\frac{1}{2}R\nabla_{\mu}\nabla_{\nu}J \nonumber\\
	&-\omega\partial_\mu\phi\partial_\nu\phi+\frac{\omega}{2}g_{\mu\nu}\partial_\sigma \phi\partial^\sigma\phi+\frac{\omega}{8}J' G\partial_\mu\phi\partial_\nu\phi =0,
	\end{align}
where $ \square \equiv \nabla_{\mu}\nabla^{\mu} $ is the d’Alembert operator, $ \nabla_{\mu} $ is the covariant derivative, $ J'\equiv \partial J/\partial X $, and $J'' \equiv \partial^2 J/\partial X^2$. In addition, $G_{\mu\nu}\equiv R_{\mu\nu}-g_{\mu\nu}R/2$ is the Einstein tensor. On the other hand,  the corresponding field equation for the scalar field $\phi$ can be defined to be
\begin{align}\label{2.4}
	\left(1-\frac{1}{8}  J' G\right)\square\phi-\frac{1}{8}G\nabla_\mu J'\nabla^\mu\phi-\frac{1}{8}J'\nabla_\mu G\nabla^\mu\phi=0.
\end{align}
In this paper, we will focus on the homogeneous and isotropic, spatially flat Friedmann-Lemaitre-Robertson-Walker (FLRW) metric given by
\begin{align}\label{2.5}
	ds^2=-dt^2+\exp[2\alpha(t)](dx^2+dy^2+dz^2),
\end{align}
where $\exp[\alpha(t)] $ is the scale factor, whose value depends only on the cosmic time $t$ due to the homogeneity of space. As a result, the corresponding definition Gauss-Bonnet term becomes as $G=24\dot{\alpha}^2\left(\dot{\alpha}^2+\ddot{\alpha}\right)$.
Consequently, the corresponding non-vanishing components of Einstein field equation now reduce to
\begin{align}\label{2.6}
	\dot{\alpha}^2&=\dot{\alpha}^3\dot{J}+\frac{\omega}{6}\dot{\phi}^2-\frac{\omega}{24}J' G\dot{\phi}^2, \\
	\label{2.7}
	2\left(1-\dot\alpha \dot J \right)\ddot{\alpha}&=-3\dot{\alpha}^2+\dot{\alpha}^2\ddot{J}+2\dot{\alpha}^3 \dot{J}-\frac{\omega}{2}\dot{\phi}^2,
	\end{align}
along with the corresponding equation of motion for the scalar field given by
\begin{align}
	\label{2.8}
	 \left(1-\frac{1}{8} J' G \right) \left(\ddot{\phi}+3\dot{\alpha} \dot{\phi}\right) = \frac{1}{8}\dot\phi \frac{d}{dt}\left (J' G \right).
	 \end{align}
It should also be noted that Eq. \eqref{2.6} is nothing but the Friedmann equation which plays as a constraint equation, while Eq. \eqref{2.7} acts as an evolution equation of scale factor $\alpha(t)$. 
%%%%%%%%%%%%%%%%%
\section{Power-law inflation}
We will seek power-law inflationary solutions \cite{Abbott:1984fp,Lucchin:1984yf,Unnikrishnan:2013vga} to the scale factor by considering the following ansatz \cite{Do:2020hjf,Do:2017qyd,Do:2011zza,Kanno:2010nr}
\begin{align}\label{2.10}
	\alpha=\zeta\log(t);~\phi=\xi\log(t)+\phi_0,
\end{align}
along with a compatible power-law coupling function given by
\begin{align}\label{2.11}
 J(X)=\lambda X^n.
\end{align} 
Here $\phi_0$, $n$, and $\lambda $ are arbitrary constants. It appears that the scale factor is now of the corresponding power-law form, $\exp[2\alpha(t)]= t^{2\zeta}$. Consequently, it turns out that $\zeta >0$ for an expanding universe,  while $\zeta > 1$ for an inflationary universe \cite{Abbott:1984fp,Lucchin:1984yf}. In many other inflationary models, $\zeta\gg 1$ seems to be a sufficient constraint for an inflationary universe in order to be consistent with the observational data of Planck, e.g., see Refs. \cite{Do:2020hjf,Unnikrishnan:2013vga,Kanno:2010nr}. As a result,  Eqs. \eqref{2.6}, \eqref{2.7}, and \eqref{2.8} can be reduced to the following set of algebraic equations,
\begin{align}\label{2.12}
		-\zeta^2+4\lambda\omega^{-1}\xi^{-2}\zeta^4+\frac{\omega\xi^2}{6}&=0,\\
		\label{2.13}
	2\zeta-3\zeta^2+4\lambda\omega^{-1}\xi^{-2}(2\zeta^3-\zeta^2)-\frac{\omega\xi^2}{2}&=0, \\
	\label{2.14}
	\left(1-3\zeta \right)\left[\omega\xi^2+12\lambda\omega^{-1}\xi^{-2}\zeta^2(\zeta^2-\zeta)\right]&=0,
\end{align} 
respectively. Here, the corresponding constraint that $ n=-1$ has been used in order to make all the field equations proportional to $t^{-2} $. In addition, it appears that $d\left (J' G \right)/dt =0$ for this power-law solution.  Interestingly, this relation implies that $G\sim (J')^{-1} \sim X^2 $ and therefore $JG \sim X$. Therefore, the coupling $JG$ will not meet any singularity issue, which ones might concern due to the negative power of $X$, in the vacuum limit with $X \to 0$.
 
\begin{figure}[!hbtp]
	\centering
	\includegraphics[scale=0.5]{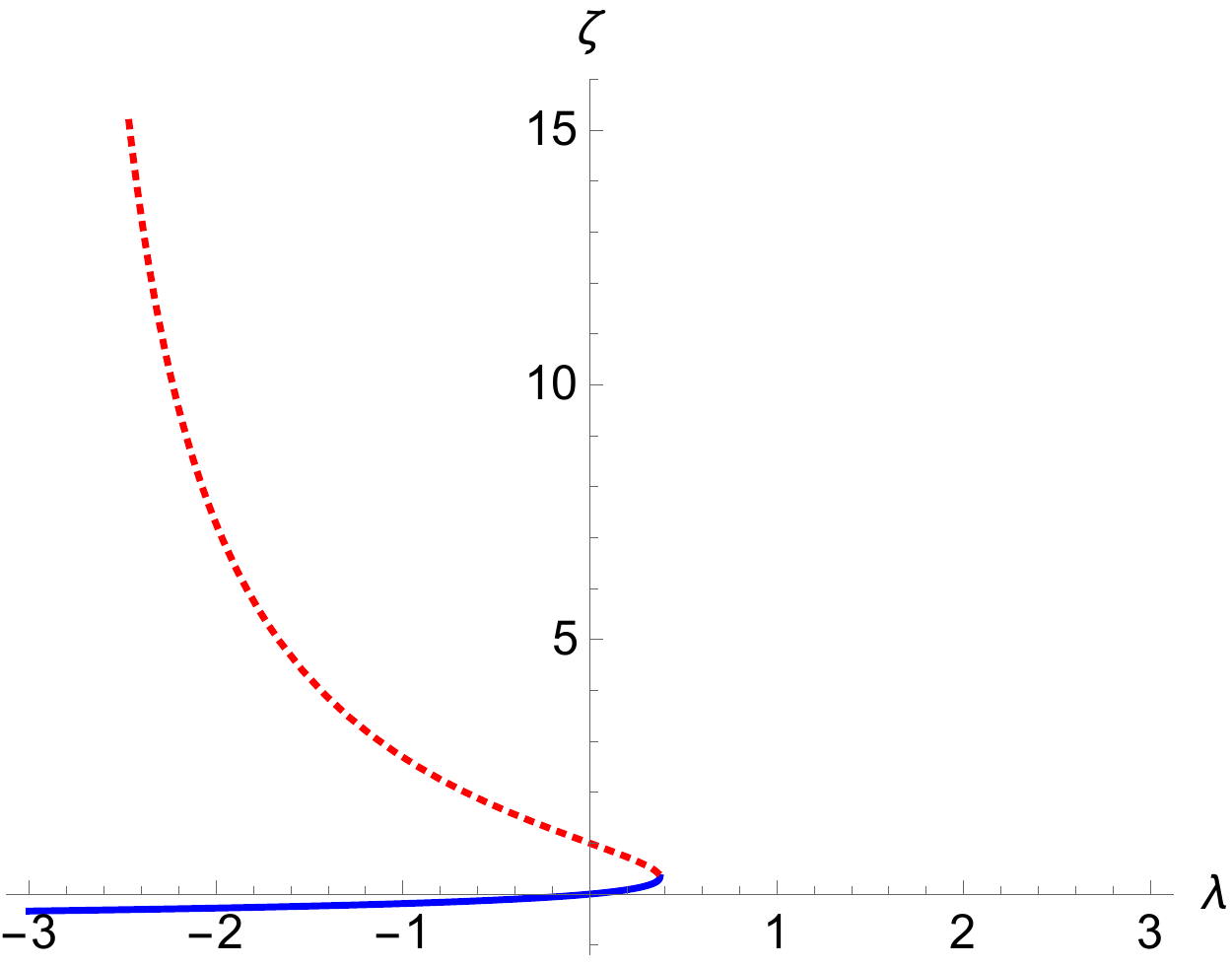}
	\caption{$ \zeta_\pm $ as a function of $ \lambda$. Here, the upper dotted red and lower solid blue curves correspond to $\zeta_+$ and $\zeta_-$, respectively.}
	\label{fig1}
\end{figure}

Now, we are going to solve these equations analytically.  As a result, we obtain from Eq. \eqref{2.14} that
\begin{align}\label{2.15}
	\omega\xi^{2}=-12\lambda\omega^{-1}\xi^{-2}\zeta^2(\zeta^2-\zeta),
\end{align}
where the trivial solution, $\zeta=1/3$, which is suitable for an expanding universe rather than an inflationary one, has been ignored.
Thanks to this relation,  Eq. \eqref{2.12} can be solved to give
\begin{align}\label{2.16}
	\omega\xi^{2}=2\lambda \zeta(\zeta+1).
\end{align}
Then, plugging this solution  into Eq \eqref{2.13} leads to an equation of $\zeta$ as
\begin{align}\label{2.17}
	(\lambda +3) \zeta ^2 +  (2 \lambda -3) \zeta +\lambda =0.
\end{align}
As a result, non-trivial solutions of $ \zeta $ can be solved to be
\begin{align}\label{2.18}
	\zeta_{\pm}=\frac{ 3-2 \lambda\pm \sqrt{3(3-8 \lambda) }}{2 (3+\lambda)}.
\end{align}
However, it turns out that only the solution, $\zeta=\zeta_+$,
is suitable for the inflationary (see Fig. \ref{fig1} for details). As a result, the constraint {\bf\color{blue} $ \zeta > 1 $} implies the following allowed region of $\lambda$ such as $-3<\lambda < 0 $. It appears that when $\lambda$ is close to $-3$, i.e., $\lambda =-3+\epsilon$ with $0<\epsilon \ll 1$, then the approximated value of $\zeta=\zeta_+$ turns out to be $\zeta \simeq {9}/{\epsilon} \gg 1$.
More interestingly, the negativity of $\lambda$ implies that  $ \omega=-1 $ according to Eq. \eqref{2.16}. Hence, $\phi$ should be phantom \cite{Caldwell:1999ew,Do:2011zza} in this model in order to have inflation. It is noted that in the usual Gauss-Bonnet inflation model having a coupling $f(\phi)G$ the scalar field should also be phantom in order to admit isotropic power-law inflation in the absence of $V(\phi)$ \cite{Guo:2009uk}. It should be noted that the power-law inflation in canonical scalar models has been shown to process a graceful exit problem \cite{Unnikrishnan:2013vga}.  Therefore, ones might ask if the power-law {\it k}GB inflation admits a graceful exit. In order to deal with this important issue, we should first note that the power-law {\it k}GB inflation is not potentially but kinetically driven, similar to the {\it k}-inflation \cite{ArmendarizPicon:1999rj,Garriga:1999vw} as well as the Galileon inflation \cite{Kobayashi:2010cm,Kobayashi:2011nu}. Therefore, $\phi$ will not oscillate after inflation because of the absence of $V(\phi)$. Furthermore, the {\it k}GB model has the shift-symmetry of $\phi$, which would prevent direct interactions between $\phi$ and standard-model contents \cite{Kobayashi:2010cm}. All these facts indicate that it seems to be very hard to have a graceful exit in a common manner within the {\it k}GB inflation. Fortunately, recent studies in Refs. \cite{Kobayashi:2010cm,BazrafshanMoghaddam:2016tdk,Hashiba:2018tbu} have pointed out that a graceful exit from the kinetically driven inflation is possible via a gravitational particle production, a.k.a. gravitational reheating \cite{Ford:1986sy,Ford:2021syk,Haque:2022kez}. Hence, it is expected that a graceful exit is also possible within the power-law {\it k}GB inflation via the gravitational reheating. A detailed investigation of this issue will be presented elsewhere.
%%%%%%%%%%%%%%%%%%
\section{Stability analysis}
We will investigate the stability of the obtained power-law solution  through a dynamical system method.  To do this, we will transform the field equations into the corresponding dynamical system of autonomous equations by introducing dimensionless dynamical variables such as 
\begin{align}\label{3.12}
	\hat{X} =\frac{\dot{\phi}}{\dot{\alpha}};~ Y=\dot{J}\dot{\alpha}.
\end{align}
As a result, we are able to define the following autonomous equations
\begin{align}\label{3.13}
	\frac{d \hat{X}}{d\alpha}&=\frac{\ddot{\phi}}{\dot{\alpha}^2}-\hat{X}\frac{\ddot{\alpha}}{\dot{\alpha}^2},\\
	\label{3.14}
	\frac{dY}{d\alpha}&=\ddot{J}+Y\frac{\ddot{\alpha}}{\dot{{\alpha}}^2},	
\end{align}
where $\alpha\equiv \int \dot\alpha dt$ acting as the dynamical time variable. As a result, the Friedmann equation \eqref{2.6} now gives
\begin{equation} \label{JG1}
 Z  = \frac{24}{\omega{\hat X}^2} \left(\frac{\omega}{6}\hat X^2 +Y-1 \right),
\end{equation} 
where $Z\equiv J' G$ an additional variable introduced for convenience. 
More interestingly, this equation implies that $Z$ should be dimensionless since $\hat X$ and $Y$ are both introduced as dimensionless dynamical variables. As a result, this requirement leads to the corresponding constraints,
\begin{equation}
\label{dyn-constraint-2}
n=-1;~Z= -\frac{96\lambda}{\hat X^4} \left(1+\frac{\ddot\alpha}{\dot\alpha^2}\right).
 \end{equation}
 As a result, we have from the field equation \eqref{2.7} that 
\begin{equation}
\frac{\ddot\alpha}{\dot\alpha^2} =\frac{1}{2\left(1-Y\right)} \left( -\frac{\omega}{2}\hat X^2 +2Y+\ddot J-3 \right),
\end{equation}
which helps us to rewrite $Z$ in Eq. \eqref{dyn-constraint-2} as 
\begin{equation} \label{JG2}
Z = -\frac{96\lambda}{\hat X^4} \left[1+\frac{1}{2\left(1-Y\right)} \left(- \frac{\omega}{2}\hat X^2 +2Y+\ddot J-3 \right) \right].
\end{equation}
As a result, comparing this equation with the equation \eqref{JG1} implies that
\begin{equation}
\ddot J =2 \left(Y-1\right) \left[ \frac{\hat X^2}{4\lambda \omega} \left(\frac{\omega}{6}\hat X^2 +Y-1 \right) +1 \right]+  \frac{\omega}{2}\hat X^2 -2Y +3 .
\end{equation}
By rewriting the field equation of scalar field, Eq. \eqref{2.8}, as
\begin{align}
	 \left(1-\frac{Z }{8} \right) \left(\ddot{\phi} +3\dot{\alpha} \dot\phi \right)=\frac{1}{8} \dot\alpha\dot{\phi} \frac{dZ}{d\alpha},
\end{align}
we are able to figure out that
\begin{align}
\frac{\ddot\phi}{\dot\alpha^2} = -3 \hat X\left(1-\frac{Z}{8} \right)-\frac{1}{8}\hat X Z \left(\frac{1}{96\lambda} \hat X^4 Z+1 \right)+\frac{Z}{8}  \frac{d\hat X}{d\alpha}+\frac{ \hat X}{8} \frac{d Z}{d\alpha},
\end{align} 
with the help of Eqs. \eqref{3.13} and \eqref{JG2}.
Thanks to these useful results, the autonomous equations \eqref{3.13} and \eqref{3.14} can now be reduced to
\begin{align}\label{3.15}
	\frac{d\hat{X}}{d\alpha}=& ~\hat X\left(1-\frac{Z}{8} \right) \left(\frac{1}{96\lambda} \hat X^4 Z-2 \right) + \frac{Z}{8}  \frac{d\hat X}{d\alpha}+\frac{\hat X}{8}  \frac{d Z}{d\alpha}  ,\\
	\label{3.16}
	\frac{dY}{d\alpha}=& ~\left(Y-2\right) \left[ \frac{\hat X^2}{4\lambda \omega} \left(\frac{\omega}{6}\hat X^2 +Y-1 \right) +1 \right] +  \frac{\omega}{2}\hat X^2 -2Y +3 ,
\end{align}
where $Z$ has been defined as a function of $\hat X$ and $Y$ in the constraint equation \eqref{JG1}. It appears that 
\begin{equation}
\frac{dZ}{d\alpha} =-\frac{24}{\omega \hat X^2} \left[\frac{2}{\hat X} (Y-1) \frac{d\hat X}{d\alpha} -\frac{dY}{d\alpha} \right].
\end{equation}
In the next steps, we will solve fixed points to these equations. Then, we will investigate the stability and attractor behavior of these fixed points.  Now, we would like to seek fixed points to the dynamical system. Mathematically, the fixed points are solutions to the following equations, $d\hat X/d\alpha =dY/d\alpha =0$. It is clear that $dZ/d\alpha =0$ for these fixed points. Consequently, we have the following equations for figuring out non-trivial fixed points $\hat X \neq 0$,
\begin{align} \label{eq-fixed-point-1}
\left(1-\frac{Z}{8} \right) \left(\frac{1}{96\lambda} \hat X^4 Z-2 \right)   &=0,\\
\label{eq-fixed-point-2}
\left(Y-2\right) \left[ \frac{\hat X^2}{4\lambda \omega} \left(\frac{\omega}{6}\hat X^2 +Y-1 \right) +1 \right] +  \frac{\omega}{2}\hat X^2 -2Y +3 &=0.
\end{align}
As a result, the first equation, i.e., Eq. \eqref{eq-fixed-point-1}, implies two possible solutions: (i) $1-Z/8=0$ or (ii) $\hat X^4 Z/(96\lambda)-2=0$. We first consider the solution (i). As a result, the solution, $Z=8$, implies, according to  the constraint equation \eqref{JG1}, that $\omega \hat X^2 = 6\left(Y-1\right)$.
As a result, the corresponding non-trivial solutions of $\hat X$ and $Y$ turn out to be
\begin{align}\label{3.18}
\hat X^2_{\pm} &= \frac{1}{\omega} \left [3\pm \sqrt{3 \left(3-8  \lambda \right) } \right],\\
Y_{\pm} &= \frac{1}{6} \left[9 \pm \sqrt{3 \left(3-8 \lambda \right)} \right],
\end{align}
here we have used a result that $\omega^2=1$. It is straightforward to see that $\hat X^2_- =\xi^2/\zeta_+^2$, where $\xi$ and $\zeta_+$ have been found previously for the power-law solution. This result indicates that the fixed point $\left(\hat X _-, Y_-\right)$ is indeed equivalent to the power-law solution. Therefore, the stability and attractor properties of the fixed point will be that of the power-law solution. It should be noted that the fixed point corresponding to the solution (ii) is equivalent to the trivial solution, $\zeta=1/3$, and therefore is not suitable for an inflationary universe. Furthermore, it will be shown later that this solution is not attractive. Hence, it is not our desired solution.
In order to see whether the fixed point $\left(\hat X_-, Y_-\right)$ is stable or not, we will perturb the dynamical system around the fixed point. For convenience, the dynamical system will be further simplified as follows
\begin{align}
	\frac{d\hat{X}}{d\alpha}=&~\frac{ \hat{X}^3}{24 \lambda }\left(\hat{X}^2+6\right)+\hat{X}, \\
	\frac{dY}{d\alpha}=&~\frac{\hat{X}^2}{24 \lambda }  \left(\hat{X}^2-6 Y+6\right) \left(Y-2\right)  -\frac{1}{2}\left(\hat{X}^2+2 Y-2\right),
\end{align}
here $ \omega=-1 $ has been used. 
As a result, perturbing these dynamical equations around the fixed point $\left(\hat X_-, Y_-\right)$ will lead to 
\begin{align}
	\frac{d\delta\hat{X}}{d\alpha}&=\frac{3-8 \lambda -\sqrt{3(3-8 \lambda) }}{2 \lambda }\delta \hat{X},\\
	\frac{d\delta Y}{d\alpha}&=\frac{3- 8 \lambda -\sqrt{3(3-8 \lambda)}}{2 \lambda }\delta Y.
\end{align}
\begin{figure}[!hbtp]
	\centering
	\includegraphics[scale=0.4]{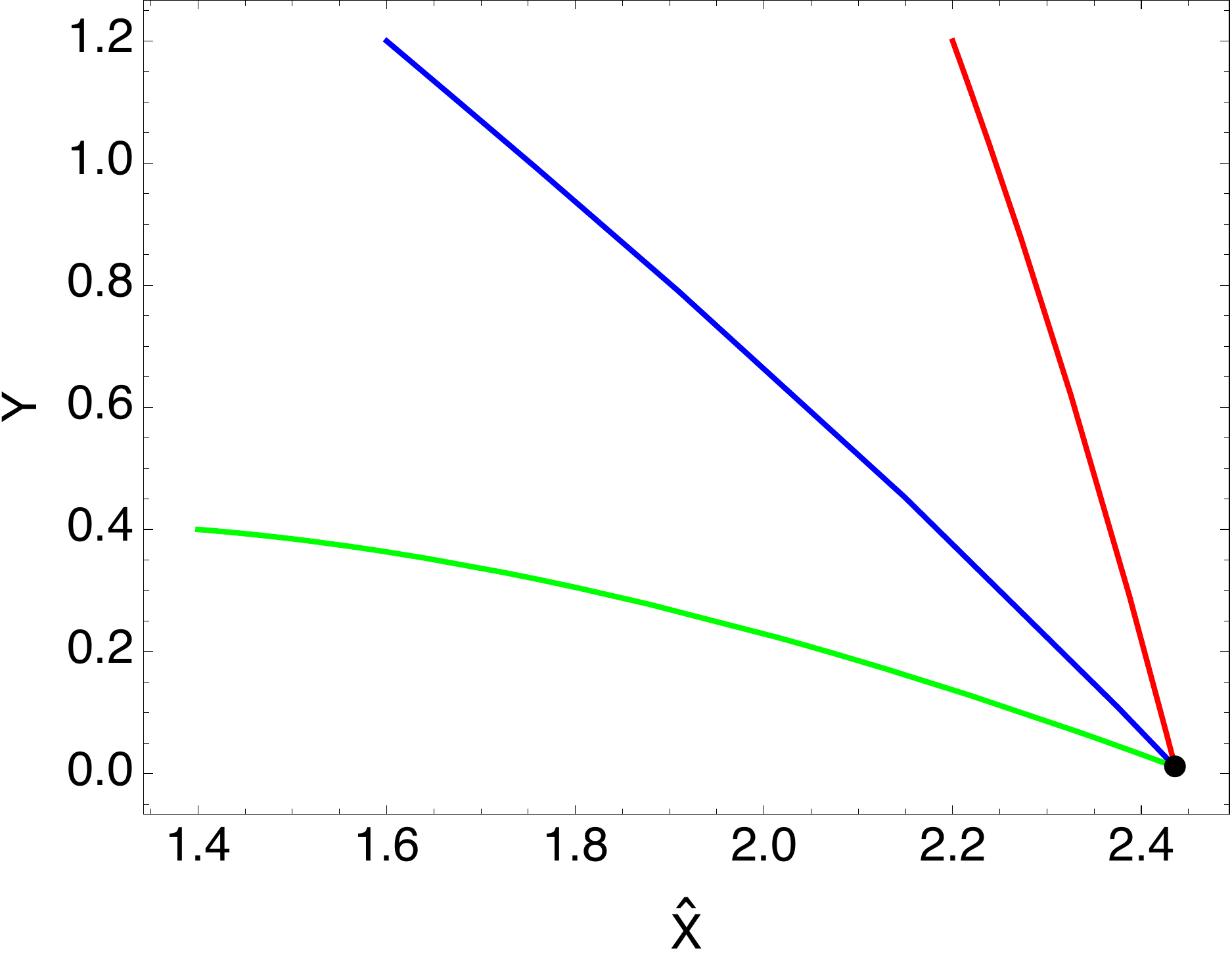}
	\caption{Attractor property of the fixed point for $\lambda=-2.95$ and $\omega=-1$. This plot clearly shows that trajectories with different initial conditions all converge to the fixed point (displayed as a black point) at late time.}
	\label{fig2}
\end{figure}
Taking exponential perturbations as $\delta \hat{X}$, $\delta Y \sim \exp[\tau \alpha]$, we will obtain the corresponding eigenvalues given by
\begin{align}
	\tau_{1,2}=\frac{3-8 \lambda -\sqrt{3(3-8 \lambda) }}{2 \lambda }.
\end{align}
Apparently, $\tau_{1,2} \simeq -3<0$ for the inflationary solution with $\lambda =-3+\epsilon $ and $0<\epsilon \ll1$. This means that the fixed point $\left(\hat X_-, Y_-\right)$ is indeed stable against field perturbations. More interestingly, we are able to numerically confirm that this fixed point is really attractive (see Fig. \ref{fig2}). However, one might ask if the {\it k}GB model admits a gradient instability in the tensor perturbations, which has been shown to exist in the Gauss-Bonnet inflation  having the usual coupling $f(\phi)G$ in the absence of  $V(\phi)$ \cite{Hikmawan:2015rze}. To deal with this issue, we now consider the tensor perturbations given by \cite{Hikmawan:2015rze,Kawai:1998ab}
\begin{equation}
ds^2 =-dt^2 +a^2(t) \left(\delta_{ij} +h_{ij} \right)dx^i dx^j,
\end{equation}
where $|h_{ij} |\ll1$. In addition, the tensor perturbations are traceless, i.e., $\delta^{ij}h_{ij}=0$, and transverse, i.e., $h_{ij,i}=0$. Consequently, the tensor perturbations carry two physical degrees of freedom, which are known as two polarizations. Interestingly, it turns out that the corresponding quadratic action for the tensor perturbations of the {\it k}GB model  is similar to that derived in Ref. \cite{Hikmawan:2015rze,Kawai:1998ab} with a replacement that $f(\phi) \to -J(X)$, 
\begin{equation}
S = \frac{1}{8} \int d^4 x a^3 \left[ (1- H\dot J ) \dot h_{ij} \dot h^{ij}-\frac{1}{a^2} (1-\ddot J ) h_{ij,k} h^{ij,k} \right],
\end{equation}
where $H\equiv \dot a/a$ being the Hubble constant.
Consequently, the corresponding squared effective speed of sound of tensor perturbations turns out to be \cite{Hikmawan:2015rze,Kawai:1998ab}
\begin{equation}
c_s^2 = \frac{1-\ddot J}{1- H\dot J}.
\end{equation}
It is clear that if $J(X)=\text{constant}$, then $c_s^2=1$ as expected.
For the power-law inflation found above, it is straightforward to define the following quantities,
$\ddot J = -{4\lambda}/{\xi^2} >0$ and $H \dot J =-{4\lambda \zeta}/{\xi^2} >0$.
Therefore, $c_s^2$ now becomes as
\begin{equation}
c_s^2 =1+\frac{2}{\zeta} \simeq 1+\frac{2}{9}\epsilon >1,
\end{equation}
with the help of Eq. \eqref{2.16}.
It is clear that $c_s^2$ in the {\it k}GB model is not negative definite, in contrast to that found in Ref. \cite{Hikmawan:2015rze}. This means that the {\it k}GB inflation  is really free from the gradient instability in the tensor perturbations. Furthermore, it should be noted that the superluminality with $c_s>1$  would not necessarily lead to  the causality violation, which is due to the appearance of the closed timelike (causal) curves, as pointed out in Refs. \cite{Bruneton:2006gf,Babichev:2007dw}. Instead, we might expect its non-trivial consequences, which might be imprinted in the CMB \cite{Mukhanov:2005bu}. It should be noted that the superluminal propagation of tensor perturbations can also be found in many other gravity models such as the Galileon inflation models \cite{Ohashi:2012wf,Bettoni:2016mij}.  To avoid the appearance of the closed time-like curves, ones could invoke the Hawking's chronology protection conjecture \cite{Hawking:1991nk} as suggested in Refs. \cite{Babichev:2007dw,Ohashi:2012wf}. Basically, this conjecture argues that the formation of the closed time-like curves could be prevented due to the significant backreaction of quantum fields happening when the time-like curves become almost closed \cite{Hawking:1991nk}. 
In addition, there is another reasonable support for the claim that the superluminality  should not be a loophole of the {\it k}GB model. In particular, ones could show, by mimicking the analysis in Ref. \cite{Bettoni:2016mij}, that the effective gravitational metric, ${\cal G}_{\mu\nu}$, which determines the causal structure of tensor propagations via the following condition ${\cal G}_{\mu\nu}dx^\mu dx^\nu =0$, of the {\it k}GB model is disformally related to the background metric $g_{\mu\nu}$, which determines the causal structure of photon propagations via the following lightcone condition $g_{\mu\nu}dx^\mu dx^\nu =0$. It is noted that both ${\cal G}_{\mu\nu}$ and $g_{\mu\nu}$ will have the same causal structure if they are conformally related to each other, i.e., ${\cal G}_{\mu\nu} = \Omega(x) g_{\mu\nu}$ \cite{Babichev:2007dw,Bettoni:2016mij}. In the {\it k}GB model, however,  it turns out, due to the disformal relation between ${\cal G}_{\mu\nu}$ and $g_{\mu\nu}$, i.e., ${\cal G}_{\mu\nu} \neq \Omega(x) g_{\mu\nu}$, as well as the superluminality $c_s>1$, that the chronology defined by ${\cal G}_{\mu\nu}$ rather than by $g_{\mu\nu}$ would be a global chronology in spacetime \cite{Bruneton:2006gf}. Therefore, the superluminality $c_s>1$ would not imply the causality violation \cite{Bruneton:2006gf}.
%%%%%%%%%%
\section{Conclusions}
We have proposed the novel {\it k}-Gauss-Bonnet gravity model, in which the scalar field $\phi$ is non-minimally kinetically coupled to the Gauss-Bonnet invariant in the absence of $V(\phi)$. As a result, the isotropic power-law inflation to this model has been solved analytically. Furthermore, this solution has been shown to be stable and attractive. More interestingly, we have pointed out that the {\it k}GB inflation  is really free from the gradient instability in the tensor perturbations. Due to the novelty of this model, however, one could ask whether its predictions, e.g., its corresponding tensor-to-scalar ratio, are consistent with the CMB observations of Planck. Furthermore, due to the existence of the third derivative of $\phi$ in the field equation \eqref{2.7} one could also ask whether the {\it k}GB model is free from the Ostrogradsky instability  \cite{Woodard:2015zca}. We will leave all these issues along with the graceful exit one mentioned above for further investigations. We hope that the {\it k}GB gravity model could provide a new approach to studies of not only the early universe but also the dark energy as well as the black hole physics.  
%%%%%%%%%%%%%%%%%%%%%%%%%%%%%%%
\begin{acknowledgments}
We would like to thank Prof. P. Kanti very much for her correspondence. We would also like to thank  Prof. A. Vikman, Prof. J. Soda, Dr. V. K. Oikonomou, and Dr. T. Paul very much for their useful comments on the previous versions of this paper.  This research is supported by the Vietnam National Foundation for Science and Technology Development (NAFOSTED) under grant number 103.01-2020.15.  
\end{acknowledgments} 
%%%%%%%%%%%%%%%%%

\end{document}